\documentclass[aps,prd,twocolumn,floatfix,nofootinbib,showpacs,superscriptaddress,tightenlines]{revtex4}
\usepackage{graphicx}
\usepackage{amsmath}
\usepackage{lipsum}
\usepackage{amssymb}
\usepackage{amsthm}
\usepackage{bbold}
\usepackage{dcolumn}    
\usepackage{epsfig}
\usepackage{graphics}
\usepackage{graphicx}
\usepackage{longtable} 
\usepackage{color}
\usepackage{epstopdf}
\usepackage{xspace}
\usepackage{cancel}
\usepackage{nicefrac}
\usepackage[colorlinks=true, pdfstartview=FitV, linkcolor=purple, citecolor= purple,urlcolor=blue]{hyperref}

\definecolor{darkgreen}{rgb}{0,0.5,0}
\definecolor{purple}{rgb}{0.5,0,0.5}
\definecolor{nblue}{rgb}{0.0,0.0,0.50}
\definecolor{scarlet}{rgb}{1.0,0.2,0}

\newcommand{\M}{\scriptsize{\hbox{M}}}

\begin{document}

\title{A multidimensional landscape of the $\eta$ and $\eta'$ mesons}

\author{L. Albino}
\affiliation{Departmento de Ciencias Integradas, Universidad de Huelva, E-21071 Huelva, Spain}
\affiliation{Departamento de Sistemas F\'isicos, Qu\'imicos y Naturales, Universidad Pablo de Olavide, E-41013 Sevilla, Spain}
\affiliation{Facultad de Ciencias Físico-Matemáticas, Universidad Autónoma de Sinaloa, Ciudad Universitaria, Culiacán, Sinaloa 80000, Mexico}
\author{K. Raya}
\affiliation{Departmento de Ciencias Integradas, Universidad de Huelva, E-21071 Huelva, Spain}
\author{R. J. Hernández-Pinto}
\affiliation{Facultad de Ciencias Físico-Matemáticas, Universidad Autónoma de Sinaloa, Ciudad Universitaria, Culiacán, Sinaloa 80000, Mexico}
\author{B. Almeida-Zamora}
\affiliation{Departamento de Sistemas F\'isicos, Qu\'imicos y Naturales, Universidad Pablo de Olavide, E-41013 Sevilla, Spain}
\affiliation{Departamento de Investigaci\'on en F\'isica, Universidad de Sonora,
Boulevard Luis Encinas J. y Rosales, 83000, Hermosillo, Sonora, Mexico}
\author{J. Segovia}
\affiliation{Departamento de Sistemas F\'isicos, Qu\'imicos y Naturales, Universidad Pablo de Olavide, E-41013 Sevilla, Spain}
\author{A. Huet}
\affiliation{Facultad de Ingenier\'ia, Universidad Aut\'onoma de Quer\'etaro, %Cerro de las Campanas s/n, Colonia Las Campanas, Centro Universitario, 
Quer\'etaro, Quer\'etaro 76010, Mexico}
\author{A. Bashir}
\affiliation{Departmento de Ciencias Integradas, Universidad de Huelva, E-21071 Huelva, Spain}
\affiliation{Instituto de F\'{i}sica y Matem\'aticas, Universidad
Michoacana de San Nicol\'as de Hidalgo, Morelia, Michoac\'an
58040, Mexico}

\date{\today}

\begin{abstract}
We employ a recently proposed form-invariant algebraic model for the quark propagator and the Bethe-Salpeter amplitude of pseudoscalar mesons to study the internal structure of $\eta$ and $\eta'$ mesons. This model facilitates the construction of the Bethe-Salpeter wavefunction, whose projection onto an appropriate flavor-basis  leads to the light-front wavefunction for convenient linear combinations of the $s \bar{s}$ and $l\bar{l}\sim(u \bar{u} +  d \bar{d})$ states. 
 Using an overlap representation, we compute the valence-quark generalized parton distributions (GPDs). The construction of the model ensures that this multidimensional quantity is determined entirely by the corresponding valence-quark distribution amplitudes. 
 Once the GPDs are constructed, we carry out a straightforward derivation of other desired physical observables such as the distribution functions and the electromagnetic form factors. We also provide explicit comparisons with available results, demonstrating that the present model offers a consistent physical picture for all ground-state pseudoscalar mesons.

\end{abstract}


\maketitle

\date{\today}

%%%%%%%%%%%%%%%%%%%%%%%%%%%%%%%%%%%%%%%%%%%%%%%%%%%%%%%%%%%%%%%
%%%%%%%%%%%%%%%%%%%%%%%%%%%%%%%%%%%%%%%%%%%%%%%%%%%%%%%%%%%%%%%
%%%%%%%%%%%%%%%%%%%%      INTRODUCTION     %%%%%%%%%%%%%%%%%%%%
%%%%%%%%%%%%%%%%%%%%%%%%%%%%%%%%%%%%%%%%%%%%%%%%%%%%%%%%%%%%%%%
%%%%%%%%%%%%%%%%%%%%%%%%%%%%%%%%%%%%%%%%%%%%%%%%%%%%%%%%%%%%%%%

\section{Introduction}
\label{SECTION Introduction}
\begin{figure}[t!]% 
%\centering
\begin{tabular}{c}
$\rule[0cm]{0.8cm}{0cm} p \rule[0cm]{0.82cm}{0cm} \pi \rule[0cm]{0.82cm}{0cm} 
K \rule[0cm]{0.82cm}{0cm} \eta \rule[0cm]{0.82cm}{0cm} \eta' \rule[0cm]{0.82cm}{0cm} \eta_c \rule[0cm]{0.82cm}{0cm} \eta_b$
\\
\hspace{-5mm}
\includegraphics[width=0.50\textwidth]{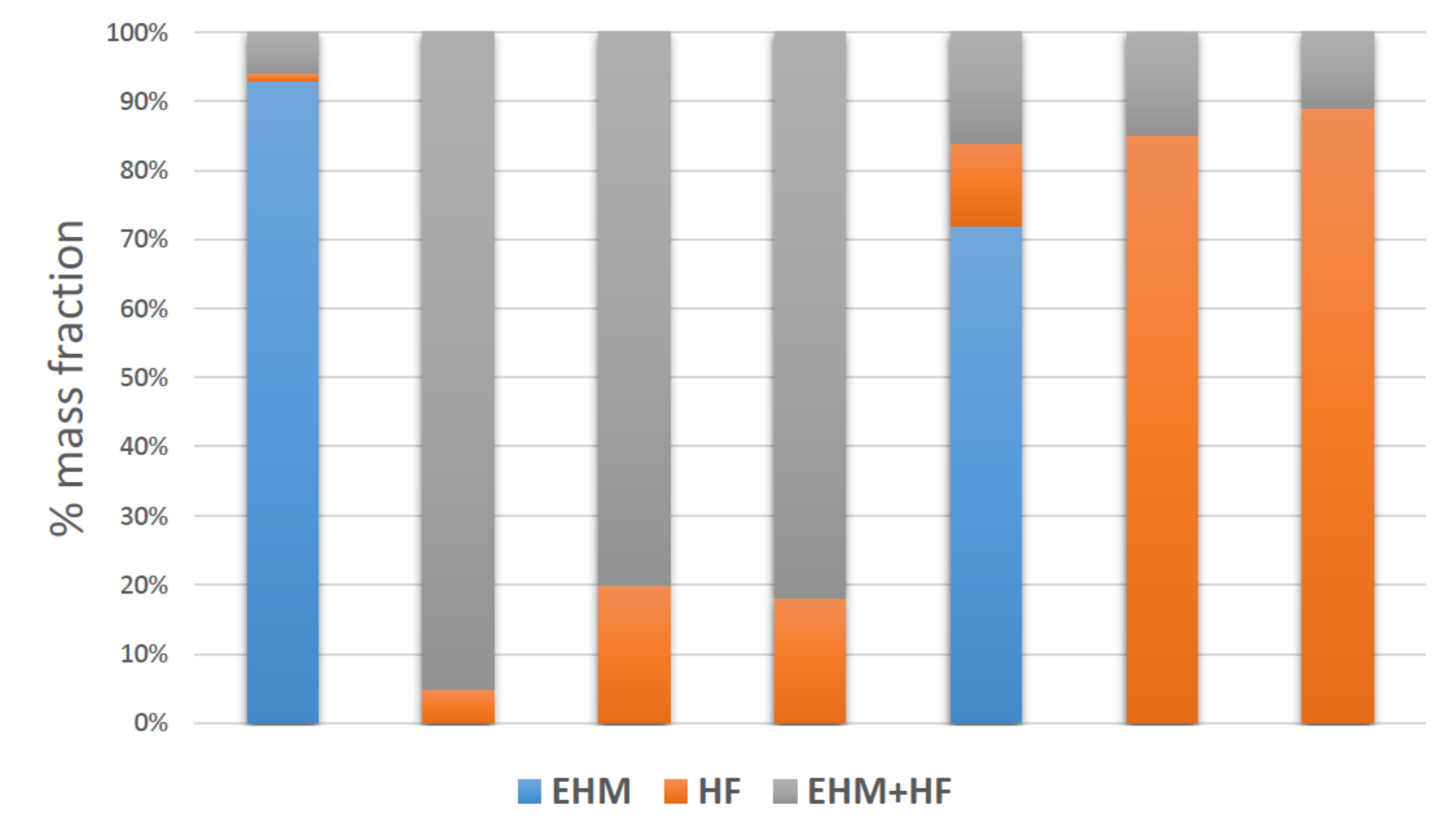}
\end{tabular}
\caption{The mass composition of different pseudoscalar mesons arises from three main sources:  emergent hadron mass (EHM), Higgs field (HF) mechanism, and the interference between between these two effects. The mass distribution within pion, kaon and $\eta$ mesons is dominated by the interplay between emergent hadron mass and the Higgs field mechanism, accounting for approximately $95$\%, $80$\% and $82$\%, of the total mass, respectively. In contrast, for the heavy-quarkonia, their mass is primarily dominated by the contribution from the Higgs field, which makes up around $85-90$\% of the total mass.  Unlike other pseudoscalars, the $\eta'$ meson derives a significant portion of its mass—approximately 70–75\%—from the emergent hadrom mass contribution, making its mass composition qualitatively comparable to that of the proton.
\label{fig:MassBuds}}
\end{figure}

From a seemingly perspicuous yet naive point of view, ground-state pseudoscalar mesons might be seen as the simplest hadrons; after all, these are merely spin-0 composite systems of negative parity which consist solely of a valence quark-antiquark pair. Expectedly and unsurprisingly, such a simplistic picture turns out to be largely inadequate and insufficient. These states carry the imprint of Nature's uncanny complexities, especially because of their intricate connection with the most dominant mass generating mechanism for the visible matter in the known universe\,\cite{Nambu:1961tp,Nambu:1961fr}. With the singular exception of the $\eta'$-meson, all other ground-state pseudoscalars can be regarded as the Nambu-Goldstone bosons arising from dynamical chiral symmetry breaking. This mechanism is a defining building block of the emergent hadronic mass and it is directly interlinked with the non-perturbative dynamics of quantum chromodynamics (QCD) in the infrared domain\,\cite{Hatsuda:1994pi,Roberts:2021nhw}. An intriguing implication of the dynamical chiral symmetry breaking is that in the absence of the mass generating mechanism by the Higgs field, the Nambu-Goldstone bosons would be identically massless. Consequently, the mass and several other structural differences observed for the pseudoscalar mesons can be attributed to the subtle interplay between the weak and the strong mass generating mechanisms. On the other hand, the non-Abelian $U_A(1)$ anomaly in QCD sets the $\eta'$ meson apart from the Nambu-Goldstone boson family and provides it with a significant mass even in the chiral limit\,\cite{Christos:1984tu,Ottnad:2025zxq}. Consequently, as displayed in Fig.~\ref{fig:MassBuds} from Refs.~\cite{Raya:2024ejx, Arrington:2021biu,Kuramashi:1994aj}, the mass composition of the $\eta'$, significantly driven by dynamical chiral symmetry breaking, markedly differs from the rest of the ground-state pseudoscalar mesons, resembling more closely that of the proton. This manifestation of the emergent hadron mass indicates that despite the $\eta$ and $\eta'$ mesons being composed of the same types of valence-quarks, and having strongly intertwined properties, a highly intricate internal dynamics is at play. These distinctive features make the mixed 
 $\eta-\eta'$ system unique and valuable for understanding the emergence of mass and its connection to the structural characteristics of hadrons\,\cite{Feldmann:1997vc,Agaev:2014wna,Ding:2018xwy}.
 %%%%%%%%%%

Nevertheless, how hadrons are formed and how their properties are dictated by the strong forces of QCD remains largely indiscernible solely from the perturbative picture provided by the QCD Lagrangian; rather, these phenomena are emergent and are inherently connected to the non-perturbative characteristics of this non-Abelian theory. It is  essential to employ a robust mathematical framework that can effectively manage the intricacies of the strong interactions, such as lattice QCD~\cite{Holligan:2025baj,Riberdy:2023awf}, and the coupled formalism of Schwinger-Dyson and Bethe-Salpeter equations~\cite{Eichmann:2016yit,Qin:2020rad}. The latter  approach connects computed hadronic observables to the Green functions of elementary degrees of freedom, namely, quarks and gluons. Moreover, the meson properties are encapsulated within the corresponding Poincaré-covariant Bethe-Salpeter wavefunction (BSWF). This framework has enabled the calculation of a plethora of physically relevant quantities related to mesons, including distribution amplitudes (DAs) and functions (DFs), electromagnetic and gravitational form factors (FFs), as well as generalized parton distributions (GPDs) and light-front wavefunctions (LFWFs) -- see \emph{e.g.}\,\cite{Roberts:2021nhw,Raya:2024ejx,Ding:2022ows} and references therein. 

Alongside these complex QCD-based computations, a series of algebraic models have been developed  which aim to streamline the calculations and facilitate the exploration of hadron structural properties without compromising the essential features of the formalism provided by the Schwinger-Dyson and Bethe-Salpeter equations (see, for example~\cite{Chang:2013pq, Raya:2015gva, Mezrag:2016hnp, Bedolla:2016yxq, Xu:2018eii,Sultan:2024mva,Raya:2021zrz}). Broadly speaking, this approach begins from coherent {\em Ans\"atze} for the quark propagator and the Bethe-Salpeter amplitude, followed by the subsequent construction of the BSWF and the LFWF\,\cite{Raya:2021zrz}. The parameters of the models are fixed by exploiting the connection between the LFWF and the experimentally accessible quantities which are deduced from it. 

The recently proposed algebraic model, Ref.~\cite{Albino:2022gzs}, generalizes several earlier attempts to a maximal form-invariant construction while preserving a plain algebraic relationship between the leading-twist LFWF and its corresponding DA. This permits the BSWF to be rigorously determined from the prior knowledge of the latter. This model has earlier been employed to study the internal structure of light pseudoscalars ($\pi-K$), heavy quarkonia and heavy-light systems~\cite{Albino:2022gzs,Higuera-Angulo:2024oui,Almeida-Zamora:2023bqb}, with explorations also conducted in the context of vector mesons~\cite{Almeida-Zamora:2023rwg}. Herein we intend to expand this series of investigations by considering the mixed $\eta-\eta'$ states. Thus, we shall apply the algebraic model outlined in the references to compute the LFWFs and GPDs of the $\eta-\eta'$ bound-states, as well as other directly computable objects, such as distribution functions, electromagnetic form factors and impact parameter space (IPS) GPDs. This allows us to deliver an in-depth perspective on the internal structure of these mesons and completes a unified picture of all ground state pseudoscalar mesons.

The manuscript is organized as follows: Sect.\,\ref{sec:AMgen} introduces the general framework of the algebraic model and explores its implications for LFWFs and GPD-related quantities. Subsequently, Sect.\,\ref{sec:MixAM} focuses on adapting this framework to the case of the mixed $\eta-\eta'$ system. The numerical results that include the analysis of DAs, DFs, GPDs, FFs and IPS-GPDs are presented in Sect.\,\ref{sec:NumRes}. Finally, Sect.\,\ref{sec:Summary} presents a summary of the present analysis and potential future perspectives.

%%%%%%%% SECTION 2
\section{An algebraic model for pseudoscalar mesons}
\label{sec:AMgen}

In this section, we summarize the key elements of the algebraic model adopted to describe pseudoscalar mesons, as introduced in Ref.~\cite{Albino:2022gzs}. Starting from an analytically tractable representation of the meson BSWF, we highlight the efficient computation of the LFWF within this model, the construction of the GPDs, and the subsequent extraction of the DF, the elastic form factors, charge radii, as well as the impact-parameter space GPDs.

\subsection{The Model and its Generalities}
\label{sec:model}

Consider a generic meson composed of a $q(\bar{h})$-flavored valence-quark (antiquark). The label $\M$  denotes quantities associated with the corresponding meson. Although our analysis focuses on the $\eta$ and $\eta'$ mesons, the general expressions presented here are valid for any pseudoscalar meson $\M$. The  BSWF (${\chi}_{\M}$) of the meson is defined as the following product of the corresponding Bethe-Salpeter amplitude, denoted as $\Gamma_{\M}$, 
and the propagators of the valence quark and antiquark, $S_{q}$ and $S_{\bar{h}}$, respectively:
\begin{eqnarray}
\label{eq:BSWF}
{\chi}_{\M} \left(k_{-}, P \right) = S_q(k) \Gamma_{\M}\left(k_{-}, P\right) S_{\bar{h}}\left(k-P\right) \,,
\end{eqnarray}
where $k_{-}=k-P/2$ and $P^2=-m_{\M}^2$, with $m_{\M}$ being the mass of the meson. Following Refs.~\cite{Albino:2022gzs,Higuera-Angulo:2024oui,Almeida-Zamora:2023bqb}, we adopt the following {\em Ansatz} for the quark (antiquark) propagator:
\begin{eqnarray}
\label{eq:quarkProp}
S_{q(\bar{h})}(k) = \left[-i\gamma \cdot k + M_{q(\bar{h})}\right] \mathrm{d}\hspace{-.5mm}\left(k^2,M_{q(\bar{h})}^2\right) 
\,,
\label{Anzats1}
\end{eqnarray}
with  $M_{q(\bar{h})}$ interpreted as the dynamically generated constituent quark (antiquark) mass and the denominator $\mathrm{d}(s,t)=(s+t)^{-1}$. For pseudoscalar mesons, the model proposes the following form for its Bethe-Salpeter amplitude:
\begin{eqnarray}
\label{eq:BSA}
n_{\M} \Gamma_{\M}(k,P) = i\gamma_5 \int_{-1}^{1} dw \, \rho(w)\left[ \hat{\mathrm{d}} \left(k_w^2 , \Lambda_w^2 \right) \right]^\nu \hspace{-.1cm}.
\label{Anzats2}
\end{eqnarray} 
This choice reflects the consideration of only the leading Bethe-Salpeter amplitude, which is sufficient for many applications, provided an appropriate spectral density, $\rho(w)$. We define $\hat{\mathrm{d}}(s,t) = t \, \mathrm{d}(s,t)$ and $ k_w = k + (w/2)P $.  $n_{\M}$ serves as a canonical normalization constant, while the exponent $\nu$ governs the asymptotic behavior of the Bethe-Salpeter amplitude, including logarithmic corrections, and may retain a degree of residual freedom depending on the chosen form of the quark propagator. Notwithstanding, to match the expectations from QCD-driven approaches\,\cite{Maris:1997tm}, our preferred choice is $\nu=1$. The last defining component of the model, $\Lambda_w^2 := \Lambda^2(w)$, encapsulates the interplay between the relevant mass scales ($M_{q,(\bar{h})},\,m_{\M}$) and the spectral density function (via $w$-dependence), as expressed by:
\begin{equation}
\Lambda^2(w) =  M_{q}^2-\frac{1}{4}\left(1-w^2\right) m_{\M}^2  +\frac{1}{2} \left( 1 -w  \right)\left(M_{\bar{h}}^2-M_q^2\right)\,.
\label{eq:Lambda}
\end{equation}
The combination of Eqs.\,\eqref{eq:BSWF}\,-\,\eqref{eq:Lambda} enables the BSWF to be conveniently cast as follows:
\begin{eqnarray}
n_{\M} \chi_{\M}(k_{-},P) \hspace{-.1cm} = \hspace{-.1cm}\mathcal{M}_{q,\bar{h}}(k,P) \hspace{-.1cm} \int_{-1}^1 \hspace{-.2cm} dw \, \tilde{\rho}^{\,\nu}_{\M}(w)\mathcal{D}^{\,\nu}_{q,\bar{h}}(k,P) \,,
\label{eq:BSWF2}
\end{eqnarray}
where the profile function, $\tilde{\rho}^{\,\nu}_{\M}(w) := \rho_{\M}(w) \, \Lambda_w^{2\nu}$, has been defined in terms of the spectral density. Moreover, the Dirac $\gamma$-matrix structure is encoded in ($p=k-P$)
\begin{eqnarray}
 \mathcal{M}_{q,\bar{h}}(k,P)  & := &  -\gamma_5 \big[ M_q\gamma \cdot P + \gamma \cdot k (M_{\bar{h}}-M_q) \nonumber \\ 
&+& \sigma_{\mu\nu}k_{\mu}P_{\nu}  - i \, (k\cdot p + M_q M_{\bar{h}}) \big] \,, \label{eq:DiracStr}
\end{eqnarray}
and $\mathcal{D}_{q,\bar{h}}^{\,\nu}(k,P)$ is a product of quadratic denominators:

\begin{eqnarray}\nonumber
\hspace{-.5cm} \mathcal{D}_{q,\bar{h}}^{\,\nu}(k,P) \hspace{-.1cm} & \equiv & \hspace{-.1cm} \mathrm{d} \left( k^2,M_q^2 \right) \mathrm{d}  \left( k_{w-1}^2,\Lambda_w^2 \right)^{\nu}  \mathrm{d}  \left( p^2,M_{\bar{h}}^2 \right) \,.\label{eq:dens}
\end{eqnarray}

Based on this structure of the BSWF, a well-defined algebraic procedure leads to the following expression:
\begin{eqnarray}
\label{eq:chits}
n_{\M}\chi_{\M} (k_{-},P)&=& \mathcal{M}_{q,\bar{h}}(k,P)  \int_0^1 d\alpha  \mathcal{F}_{\M}(\alpha,\sigma^{\nu+2})\,,
% \\  
% \mathcal{F}_{\M}(\alpha,\sigma^{\nu+2})&=&\nu(\nu+1)
%  \Big[ \int_{-1}^{1-2\alpha} dw 
% \int_{ \frac{2\alpha}{w-1} + 1}^1 d\beta \label{BSE2.1} \\
% && \hspace{-0.7 cm} + \int_{1-2\alpha}^1 dw \int_{ \frac{ 2\alpha+(w-1)}{w+1 } }^1 d\beta \Big] \frac{(1-\beta)^{\nu-1} \tilde{\rho}_{\M}^{\,\nu}(w) }{ \sigma^{\nu+2} } \,,\nonumber
\end{eqnarray}
where $\sigma =  (k-\alpha P)^2 + \Lambda^2_{1-2\alpha}$, and $\alpha$ is the usual Feynman parameter. Moreover, 
% since only the expression  $(1-\beta)^{\nu-1}$ depends on $\beta$, the integration on this variable can be performed directly, leading to:
the convenient representation chosen for $\Lambda^2(w)$, as given in Eq.\,\eqref{eq:Lambda}, entails
\begin{eqnarray}
    \label{eq:FM1}
    \mathcal{F}_{\M}(\alpha,\sigma^{\nu+2})&=&2^\nu(\nu+1)\Big[ \int_{-1}^{1-2\alpha} dw \left(\frac{\alpha}{1-w}\right)^\nu\\
    &+&\int_{1-2\alpha}^1 dw \left(\frac{1-\alpha}{1+w}\right)^\nu\Big] \frac{\tilde{\rho}_{\M}^{\,\nu}(w)}{\sigma^{\nu+2}}\;.\nonumber
\end{eqnarray} 
It is important to stress the fact that, in order to prevent the emergence of poles on the real axis, it is crucial to ensure the positivity of $\Lambda^2(w)$. This condition is satisfied as long as:
\begin{eqnarray}
|M_{\bar{h}}-M_q| \leq m_{\M} \leq M_{\bar{h}}+M_q \,.
\label{eq:positivity}
\end{eqnarray}
It is worth noting that the leftmost side of the inequality is trivially satisfied for isospin-symmetric systems (where $M_{q}=M_{\bar{h}}$) and, indeed, holds for any realistic ground-state bound systems\,\cite{Raya:2019dnh,Xu:2024vkn,Sultan:2018tet}. The remaining condition of Eq.\,\eqref{eq:positivity} does not pose a problem for Nambu–Goldstone modes, owing to the relatively small values of $m_{\M}$. However, for other mesons, a careful selection of constituent quark masses is essential, or an alternative approach must be considered. In the following subsection, we show how the Nakanishi integral representation for the BSWF, Eqs.~\eqref{eq:chits} and \eqref{eq:FM1}, yields an analytical relation between the meson LFWF and the DA. 
% Having discussed the construction of a BSWF for a generic pseudoscalar meson, we now proceed to adapt it for the mixed $\eta-\eta'$ system.

\subsection{Light-front wavefunctions}
\label{sec:LFWF}
The projection of the meson BSWF onto the light-front results in the associated LFWF. It yields the leading-twist expression for a quark $q$ within a generic pseudoscalar meson $\M$:
\begin{equation}
\psi_{\M}^q \left( x,k_{\perp}^2 \right) = \text{tr}_{\text{CD}} \int_{ dk_{\parallel} }\delta_n^x(k_{\M}) \gamma_5 \gamma \cdot  n \, \chi_{\M}(k_{-},P) \,,
\label{eq:LFWF}
\end{equation}
where $\delta_n^x(k_{\M})=\delta( n  \cdot  k - x \, n \cdot  P )$; $ n $ is a light-like four-vector, such that $ n^2 = 0 $ and $ n\cdot P = -m_{\M} $; $x$ corresponds to the light-front momentum fraction carried by the active quark. The symbol $\text{tr}_{\text{CD}}$ refers to the trace over color and Dirac indices. The compact and convenient notation $\int_{dk_\parallel} := \int \frac{d^2 k_\parallel}{\pi}$ has been employed and the 4-momentum integral is defined as usual:
\begin{eqnarray}
\int \frac{d^4k}{(2\pi)^4} = \left[ \frac{1}{16 \pi^3} \int d^2 k_\perp \right] \left[ \frac{1}{\pi} \int d^2 k_\parallel \right]\,.
\end{eqnarray}
It turns out relatively more convenient to consider the Mellin moments of the LFWF:
\begin{eqnarray}\label{eq:MellinMoments}
\langle x^m \rangle_{ \psi_{\M}^q } &=& \int_0^1 dx \, x^m \, \psi_{\M}^q \left( x,k_{\perp}^2 \right) \\
&=& \text{tr}_{\text{CD}} \frac{1}{n\cdot P} \int_{dk_{\parallel}} \left[\frac{n\cdot k}{n\cdot P} \right]^m  \gamma_5 \gamma \cdot n \chi_{\M}(k_{-},P)\,.\nonumber
\end{eqnarray}
When adopting the BSWF representation introduced in Sec.\,\ref{sec:model}, a clear sequence of straightforward algebraic steps yields:
\begin{eqnarray}
\langle x^m \rangle_{ \psi_{\M} ^q} &=& \int_0^1 d\alpha \alpha^m \left[\frac{12}{n_{\M}} \frac{\mathcal{Y}_{\M}(\alpha,\sigma_\perp^{\nu+1})}{\nu+1}\right]\,,
\label{eq:LFWFmoms}\\
\mathcal{Y}_{\M}(\alpha,\sigma_\perp^{\nu+1})&=&\mathcal{F}_{\M}(\alpha,\sigma_\perp^{\nu+1})(\alpha M_{\bar{h}}+(1-\alpha)M_q)\,,\nonumber
\end{eqnarray}
where $\sigma_\perp = k_\perp^2+ \Lambda_{1-2\alpha}^2$. Moreover, the uniqueness property of the Mellin moments, Eqs.\,\eqref{eq:LFWF} and\,\eqref{eq:MellinMoments}, establishes a connection between the Feynman parameter $\alpha$ and the momentum fraction $x$, allowing for a straightforward identification of the LFWF:
\begin{equation}
    \label{eq:LFWFgood}
\psi_{\M}^q(x,k_\perp^2)=\left[\frac{12}{n_{\M}} \frac{\mathcal{Y}_{\M}(x,\sigma_\perp^{\nu+1})}{\nu+1}\right]\,.
\end{equation}
Integrating over $k_\perp$ yields the DA:
\begin{eqnarray}
f_{\M}\varphi_{\M}^q(x) = \frac{ 1}{16\pi^3 } \int d^2 k_{\perp} \psi_{\M}^q \left( x,k_{\perp}^2 \right) \,,
\label{eq:PDAdefinition}
\end{eqnarray}
where $\varphi_{\M}^q(x)$ is unit normalized and $f_{\M}$ denotes the corresponding weak leptonic decay constant. As might be inferred from Eqs.\,\eqref{eq:FM1} and\,\eqref{eq:LFWFmoms}, the $k_\perp$ dependence of $\psi_{\M}^q(x,k_\perp^2)$ is trivial and the integration follows directly. An advantageous and desirable outcome is the algebraic relation below:
\begin{equation}
    \label{eq:LFWFPDArel}
    \psi_{\M}^q(x,k_\perp^2) = 16\pi^2 f_{\M}\frac{\nu \Lambda_{1-2x}^{2\nu}}{(k_\perp^2+\Lambda_{1-2x}^2)^{\nu+1}}\varphi_{\M}^q(x)\,.
\end{equation}
Many crucial implications of this simple algebraic connection between the LFWF and the PDA are discussed in Refs.\,\cite{Albino:2022gzs,Higuera-Angulo:2024oui,Almeida-Zamora:2023bqb}. Here, we emphasize again that with the prior knowledge of the DA, there is no need to construct the spectral density $\rho_{\M}(w)$ in any way to derive the LFWF, albeit the connection between $\rho_{\M}(w)$ and $\varphi_{\M}^q(x)$ is well-known and direct\,\cite{Albino:2022gzs}. Only the relevant mass scales, namely $m_{\M}$ and $\,M_{q,\bar{h}}$, need to be determined. 

It should be pointed out that the intrinsic resolution scale $\zeta$, at which the LFWF is defined, remains unspecified. Being a leading-twist distribution, it turns out natural to set it as the so-called hadronic scale $\zeta_H$, where the hadron's properties can be adequately captured by the dressed 
valence-quark degrees of freedom\,\cite{Raya:2021zrz}. Therefore, we expect $\zeta_H$ to be of the order of $\zeta_H \sim M_u \approx 0.3 - 0.4$ GeV\,\cite{Lu:2023yna,Yin:2023dbw,Xu:2023bwv}. We shall assume this scale from now on.

Another feature of Eq.\,\eqref{eq:LFWFPDArel} is that the $x-k_\perp$ correlations are determined by the $\Lambda_{1-2x}^2$ term, which explicitly takes the form:
\begin{equation}
    \Lambda_{1-2x}^2 = M_q^2 + x\,(M_{\bar{h}}^2-M_q^2) -x(1-x)m_{\M}^2\,.
\end{equation}
It encapsulates the fact that these correlations are driven by the extent of chiral $M_q$ and flavor $(M_{\bar{h}}^2-M_q^2)$  symmetry breaking. In the chiral limit, $\Lambda_{1-2x}^2=M_q^2$, implying
\begin{eqnarray}
\label{eq:LFWFPDAfrac2}
    \psi_{\M}^q(x,k_\perp^2)&=& \tilde{\varphi}_{\M}^q(k_\perp^2)\,   \varphi_{\M}^q(x)\;, \\
    \label{eq:LFWFPDAfrac} 
    \tilde{\varphi}_{\M}^q(k_\perp^2) &:=& 16\pi^2 f_{\M}\frac{\nu M_q^{2\nu}}{(k_\perp^2+M_q^2)^{\nu+1}}\,,
\end{eqnarray}
which  recovers the chiral limit results from Refs.\,\cite{Mezrag:2016hnp,Chouika:2017rzs,Raya:2021zrz}. Note that this corresponds to a separable Ansatz.

Finally, it might be mentioned that to obtain all antiquark related quantities, it is sufficient to replace 
\begin{equation}
    \varphi_{\M}^q(x)\to \varphi_{\M}^{\bar{h}}(x)=\varphi_{\M}^q(1-x)\,,
\end{equation}
and exchange the constituent quark masses $M_{\bar{h}} \leftrightarrow M_q $. We are now set to compute the GPDs.

\subsection{Generalized parton distributions}
\label{sec:GPD}
The valence-quark GPD can be obtained from the overlap representation of the LFWF~\cite{Diehl:2003ny}:
\begin{eqnarray}
\nonumber
 H_{\M}^q(x,\xi,\Delta^2) \hspace{-0.1 cm} &=& \hspace{-0.1 cm} \int \frac{d^2k_{\perp}}{16\pi^3} \psi_{\M}^{q*}  \left( x^-, (\mathbf{k}_\perp^{-})^2\right) \psi_{\M}^q  \left( x^+, (\mathbf{k}^{+}_\perp)^2 \right) \,, \\
\hspace{0.2 cm} x^{\pm}&=&\frac{x\pm \xi}{1\pm \xi}\hspace{0.4 cm} , \hspace{0.4 cm} \mathbf{k}_\perp^\pm=k_\perp \mp \frac{\Delta_\perp}{2}\frac{1-x}{1\pm \xi}\,.
\label{eq:GPDdefinition}
\end{eqnarray}
As before, $x$ represents the light-front momentum fraction carried by the struck quark, and $\Delta^2$ denotes the momentum squared transferred from the probe to the struck hadron. In this context, its transverse component squared is $\Delta_\perp^2 = \Delta^2(1-\xi^2) -4\xi^2 m_{\text{M}}^2$, while $\xi$,  referred to as the skewness, is the longitudinal momentum fraction transferred. Both $x$ and $\xi$ have support on $[-1,1]$, but the overlap representation is only valid in the DGLAP region $|x| > \xi$\,\cite{Chouika:2017rzs,DallOlio:2024vjv}. Exploring this domain is sufficient for several quantities of interest such as the distribution functions, electromagnetic form factors, and the impact parameter space GPDs.

Considering the LFWF representation previously derived in Eq.\,\eqref{eq:LFWFPDArel}, it follows directly that the GPD takes the following form:
\begin{equation}
\label{eq:defGPDgood}
    H_{\M}^q(x,\xi,\Delta^2)=\mathcal{N}_H \left[\frac{\varphi_{\M}^q(x^+)\varphi_{\M}^q(x^-)}{\Lambda_{1-2x^+} \Lambda_{1-2x^-}}\right] \Phi_{\M}^q(x,\xi,\Delta^2)\,, 
\end{equation}
where the normalization constant $\mathcal{N}_H$ ensures unit normalization in the forward limit, where the associated DF is defined, \emph{i.e.}:
\begin{eqnarray}
\label{eq:defPDF}
&&  \hspace{-5mm}  q_{\M}(x) := H_{\M}^q(x,0,0) = \mathcal{N}_H \frac{[\varphi^q_{\M}(x)]^2}{\Lambda_{1-2x}^2}\,,\\
&&  \hspace{-7mm}  \int_0^1 q_{\M}(x)\,dx = \int_0^1 \mathcal{N}_H \frac{[\varphi^q_{\M}(x)]^2}{\Lambda_{1-2x}^2}\,dx=1\,.
\end{eqnarray}
Notice that the closed form of the relation between DA and DF, \emph{i.e.}, Eq.\,\eqref{eq:defPDF},  is valid only at $\zeta_H$. Moreover, one can readily deduce from Eq.\,\eqref{eq:defGPDgood} that $\Phi_{\M}^q$ drives the $\Delta^2$ evolution of the GPDs. This function reads as:
\begin{equation}
\label{eq:defPhi}
    \Phi_{\M}^q(x,\xi,\Delta^2) = \mathcal{N}_{\Phi} \Lambda_{1-2x^+}^{2\nu+1} \Lambda_{1-2x^-}^{2\nu+1}\int_0^1 du \frac{u^\nu(1-u)^\nu}{[\mathbb{M}^2(u)]^{2\nu+1}}\,,
\end{equation}
where
\begin{eqnarray}
\mathbb{M}^2(u) &=& z\, u(1-u)+ u (\Lambda^2_{1-2x^+}-\Lambda^2_{1-2x^-})+\Lambda^2_{1-2x^-}\,,\nonumber\\ 
\mathcal{N}_\Phi &=& \frac{\Gamma(2\nu+2)}{\Gamma^2(\nu+1)}\,,\,z:=\frac{(1-x)^2}{(1-\xi^2)^2}\Delta_\perp^2\,.\label{eq:defM}
\end{eqnarray}
It is also noteworthy that for specific values of $\nu > -1$, the integral in Eq.\,\eqref{eq:defPhi} can be evaluated analytically. This highlights the practicality of the algebraic model. Focusing on the chiral limit $(c.l.)$, one has
\begin{equation}
    \mathbb{M}^2(u) \overset{c.l.}{=} z\,u(1-u)+M_q^2\,,
\end{equation}
so that $\Phi_{\M}^q$ is fully expressed as a function of the single kinematic variable  $z$. The GPD then reduces to
\begin{eqnarray}\nonumber
    H_{\M}^q(x,\xi,\Delta^2) &=&  \mathcal{N}_H \left[\frac{\varphi_{\M}^q(x^+)\varphi_{\M}^q(x^-)}{M_q^2}\right] \Phi_{\M}^q(z)\\
    &=&\mathcal{N}_H \sqrt{q_{\M}(x^+)q_{\M}(x^-)}\, \Phi_{\M}^q(z)\,.
    \label{eq:GPDfac}
\end{eqnarray}
The second line is merely a consequence of the definition of the DF, Eq.\,\eqref{eq:defPDF}, whose chiral limit evaluation yields:
\begin{equation}
    \label{eq:PDFfac}
    q_{\M}(x;\zeta_H) \propto [\varphi_{\M}^q(x;\zeta_H)]^2\,.
\end{equation}
Notice that we explicitly reinstate the model scale $\zeta_H$ here which, as mentioned before, is implicitly employed throughout this manuscript. The reason is that DA and DF evolve in a different manner. Thus, Eq.\,(\ref{eq:PDFfac}) would not be valid beyond the model scale. 
Moreover, the relationship between the distributions, established by Eqs.\,\eqref{eq:GPDfac} and\,\eqref{eq:PDFfac}, is a defining feature of systems characterized by a factorized LFWF, Eq.\,\eqref{eq:LFWFPDAfrac2}. For this family of Ans\"atze, regardless of the specific shape of $\Phi_{\M}^q(z)$ and $\tilde{\varphi}_{\M}^q(k_\perp^2)$, an array of additional algebraic outcomes follow\,\cite{Raya:2021zrz,Raya:2024glv}.

\subsection{Electromagnetic form factors}
\label{sec:EFFs}

Electromagnetic form factors are the zeroth Mellin moment of the GPD. Explicitly, the contribution of the valence-quark $q$ to the FF is given by:
\begin{equation}
\label{eq:EFFq}
F_{\M}^q(\Delta^2)=\int_0^1dx\,H_{\M}^q(x,\xi=0,\Delta^2)\,,
\end{equation}
so that the complete result follows from adding up the individual components:
\begin{equation}
\label{eq:EFFtot}
    F_{\M}(\Delta^2) \equiv e_q F_{\M}^q(\Delta^2)+e_{\bar{h}}F_{\M}^{\bar{h}}(\Delta^2)\,;
\end{equation}
clearly, $e_q$ and $e_{\bar{h}}$ are the valence-quark charges in units of the positron's electric charge. The integration domain in Eq.\,\eqref{eq:EFFq} reflects our choice of skewless GPD,  $\xi=0$, since the electromagnetic form factor is independent of this variable\,\cite{Mezrag:2023nkp}. The associated charge radius $r_{\M}$ is defined as usual:
\begin{eqnarray}
(r_{\M})^2&=& e_q (r_{\M}^{q})^2+e_{\bar{h}}(r_{\M}^{\bar{h}})^2\,,\\
    (r_{\M}^{q})^2&:=&-6 \frac{\partial F_{\M}^q(\Delta^2)}{\partial \Delta^2} \Bigg|_{\Delta^2=0}\,.\label{eq:chargeAn}
\end{eqnarray}
Setting $\xi \hspace{-.8mm}=\hspace{-.8mm}0$ and expanding Eq.\,\eqref{eq:defGPDgood} around $\Delta^2 \hspace{-1mm}=\hspace{-.6mm} \Delta_\perp^2 \hspace{-1mm}\approx\hspace{-.6mm} 0$:
\begin{eqnarray}
\nonumber
H_{\M}^q(x,0,\Delta^2) &\overset{\Delta^2\to 0}{\approx}&  \mathcal{N}_H\frac{[\varphi_{\M}^{q}(x)]^2}{\Lambda_{1-2x}^2} \Big[1 - \mathfrak{L}_{\M}^q(x) \Delta^2 + ... \Big] \,,\\
\label{eq:GPDlowQ2}
    \mathfrak{L}_{\M}^q(x) &:=& \frac{(1+\nu)(1+2\nu)}{2(3+2\nu)}\frac{(1-x)^2}{\Lambda_{1-2x}^2} \,.
\end{eqnarray}
It follows that
\begin{equation}
\label{eq:crquark}
    (r_{\M}^q)^2=6 \, \int_0^1 dx\; \mathfrak{L}_{\M}^q(x)q_{\M}(x) \,.
\end{equation}
The antiquark component is obtained in an analogous way. In quarkonium systems $F_{\M}=0$ and $r_{\M}=0$, since we have $e_q+e_{\bar{h}}=0$. Therefore, when presenting electromagnetic FFs and charge radii, we shall only consider the individual flavor contribution ($F_{\M}\to F_{\M}^q$ and $r_{\M}\to r_{\M}^q$). Moreover, if the charge radius is known experimentally, it can be employed to tune the mass scales required to define the algebraic model aligned with phenomenology.

\subsection{Impact parameter space GPDs}
The impact parameter space GPD can be obtained in a straightforward manner by carrying out the Fourier transform of the zero-skewness GPD \cite{Burkardt:2000za}:
\begin{eqnarray}
\textbf{q}_{\M}(x,|b_\perp|)= \int_0^{\infty}\frac{d\Delta}{2\pi }\Delta J_0 (|b_{\perp}| \Delta) H_{\M}^q(x,0,\Delta^2) \,,
\end{eqnarray} 
where $J_0$ denotes the cylindrical Bessel function. Naturally, this distribution represents 
the probability density of finding a parton with momentum fraction $x$ at a transverse distance $|b_{\perp}|$ from the center of transverse momentum of the meson under study. 

As suggested in Ref.\,\cite{Albino:2022gzs}, we consider the following approximation for the zero-skewness GPD:
\begin{equation}
\label{eq:GPDexp}
    H_{\M}^q(x,0,\Delta^2)= q_{\M}(x) \;\text{exp}[- \mathfrak{L}_{\M}^q(x) \Delta^2]\,.
\end{equation}
This particular form provides additional algebraic benefits that enhance those of the present model\,\cite{Raya:2024glv}. In particular, it allows for a straightforward derivation of the IPS-GPD (here $\mathcal{I}_{\M}^q(x,|b_\perp|):=2\pi|b_\perp|\textbf{q}_{\M}(x,|b_\perp|)$):
\begin{equation}
    \label{eq:IPSdef}
    \mathcal{I}_{\M}^q(x,|b_\perp|)=\frac{q_{\M}(x)}{2 \mathfrak{L}_{\M}^q(x)}|b_\perp|\,\text{exp}\left[-\frac{|b_\perp|^2}{4 \mathfrak{L}_{\M}^q(x)}\right]\,.
\end{equation}
As is the case with all other distributions, the form of $\mathcal{I}_{\M}^q(x, |b_\perp|)$ is expected to reflect the influence of mass generation mechanisms. This connection arises naturally because the IPS-GPD is explicitly built from the parton distribution function which encodes the longitudinal momentum dependence. Consequently, the broadness of $\mathcal{I}_{\M}^q$ in transverse space and the height of its peak directly trace the behavior of $q_{\M}(x)$, providing insight into the spatial distribution of partons and its correlation with dynamical mass generation~\cite{Raya:2021zrz,Raya:2024glv}.

\section{The $\eta-\eta'$ case}
\label{sec:MixAM}
For the description of the $\eta-\eta'$ mesons, we shall consider the $SU(2)$ flavor symmetry, i.e., the isospin symmetry which entails $M_u=M_d \equiv M_l$. This approximation results in the complete decoupling of the neutral pion from the intertwined system of $\eta-\eta'$ mesons, which features $s\bar{s}$ components~\cite{Bhagwat:2007ha}. We follow Ref.\,\cite{Ding:2018xwy} and express these physical states in a generic flavor basis:
\begin{equation}
\label{eq:chisl}
\chi_{\eta,\eta'}(q,P) = \mathcal{T}_l \chi_l^{\eta,\eta'}(q,P) + \mathcal{T}_s \chi_s^{\eta,\eta'}(q,P)\,,
\end{equation}
with $\mathcal{T}_l=\text{diag}(1/2,1/2,0)$ and $\mathcal{T}_s=\text{diag}(0,0,1/\sqrt{2})$. Consequently, the mixed $\eta-\eta'$ system is fully characterized by four BSWFs: $\chi_{l,s}^{\eta,\eta'}(q,P)$. 

We choose to work with the single-angle mixing scheme (SA-MS) approximation~\cite{Feldmann:1998sh,Feldmann:1998vh}:
\begin{eqnarray}
\label{eq:saBasis0}
\left(\begin{array}{cc}
     \eta  \\
     \eta' 
\end{array}\right)  =
U(\theta) \left(\begin{array}{cc}
     (u\bar{u}+d\bar{d})/\sqrt{2}  \\
     s\bar{s} 
\end{array}\right) =
U(\theta) \left(\begin{array}{cc}
     \eta_l  \\
     \eta_s 
\end{array}\right)\,,
\end{eqnarray}
where $\theta$ is the mixing angle and 
\begin{eqnarray}
    U(\theta) &:=&
\left(\begin{array}{cc}
     \cos \theta & -\sin \theta \\
     \sin \theta & \quad \cos \theta 
\end{array}\right)\,.
\end{eqnarray}
Here $\eta_l$ and $\eta_s$ denote pure $l\bar{l} \sim(u \bar{u} +  d \bar{d})$ and $s\bar{s}$ orthonormal states, respectively. Eq.\,\eqref{eq:saBasis0} implies the following mixing for the corresponding BSWFs:
\begin{eqnarray}
\label{eq:saBasis}
\left(\begin{array}{cc}
     \chi_{\eta}(q,P)  \\
     \chi_{\eta'}(q,P) 
\end{array}\right) &=& U(\theta) \left(\begin{array}{cc}
     \chi_{\eta_l}(q,P)  \\
     \chi_{\eta_s}(q,P) 
\end{array}\right)\,.
\end{eqnarray}
Clearly, $\theta=0$ implies the physical $\eta-\eta'$ mesons would become identical to the $\eta_l$ and $\eta_s$ states, respectively, and $\eta_l$ would exhibit pion-like characteristics. In the chiral limit, this would correspond identically to a massless pion. In the realistic scenario, $\theta \neq 0$ provides a measure of the appropriate degree of mixing between the $\eta_l$ and $\eta_s$ states to produce the observed 
physical states $\eta$ and $\eta'$, indicating a measure of the flavor content in the physical $\eta-\eta'$ mesons.

One of the key advantages of representing the $\eta-\eta'$ mesons in the SA-MS is that extra subtleties are avoided concerning the positivity condition in Eq.\,\eqref{eq:positivity}. Note that on adopting the representation from Eq.\,\eqref{eq:chisl}, the substantial mass of the $\eta'$ meson ($m_{\eta'}=0.96$ GeV) would expectedly lead to a violation of the positivity condition, as realistic values for $M_{u,s}$ typically range from $0.3$ to $0.6$ GeV\,\cite{Sultan:2018tet}. SA-MS circumvents this problem. Assuming SA-MS is a fair approximation, one can readily establish a mapping between bases:
\begin{eqnarray}
 \left(\begin{array}{cc}
     \chi_{\eta_l} & 0  \\
     0 & \chi_{\eta_s}
\end{array}\right) = U^{-1}(\theta) \left(\begin{array}{cc}
     \chi_{\eta}^l & \chi_{\eta}^s  \\
     \chi_{\eta'}^l & \chi_{\eta'}^s
\end{array}\right) \,,\label{eq:mapping}
\end{eqnarray}
which translates into
\begin{eqnarray}
 \left(\begin{array}{cc}
     f_{\eta_l} & 0  \\
     0 & f_{\eta_s}
\end{array}\right) = U^{-1}(\theta) \left(\begin{array}{cc}
     f_{\eta}^l & f_{\eta}^s  \\
     f_{\eta'}^l & f_{\eta'}^s
\end{array}\right) \,.\label{eq:mapping2}
\end{eqnarray}
Evidently, $f_{\M}^{(q)}$ denotes the corresponding decay constants of the $\M$-meson. Equipped with the pseudoscalar meson BSWF and the mixing scheme, we can derive a coherent and adequate representation of the corresponding LFWF. 

Setting $\zeta_H$ as the resolution scale, the Fock-space expansion of the $\eta-\eta'$ bound-states follows directly from Eq.\,\eqref{eq:saBasis0}. Therefore, the mixed system is completely determined by the mixing angle $\theta$ and the LFWFs of the states $\eta_l$ and $\eta_s$. Naturally, these retain the same form as given in Eq.~\eqref{eq:LFWFPDArel}:
\begin{equation}
    \label{eq:LFWFPDArel2}
    \psi_{\eta_l[\eta_s]}^{l[s]}(x,k_\perp^2) = 16\pi^2 f_{\eta_l,\eta_s}\frac{\nu \Lambda_{1-2x}^{2\nu}}{(k_\perp^2+\Lambda_{1-2x}^2)^{\nu+1}}\varphi_{\eta_l[\eta_s]}^{l[s]}(x)\;.
\end{equation}
The input of $\varphi_{\eta_l[\eta_s]}^{l[s]}(x)$ is thus required. However, realistic DAs regarding the $\eta-\eta'$ system have only been reported in the generic flavor basis of Eq.\,\eqref{eq:chisl}, Ref.\,\cite{Ding:2018xwy}. Therefore, we consider the mapping in Eq.\,\eqref{eq:mapping} to produce the following relations among LFWFs:
\begin{eqnarray}
\label{eq:MapLFWF}
\begin{array}{cc}
  \hspace{-4mm}   \psi_{\eta_l}^l (x,k^2_{\perp}) =  \cos \theta \, \psi^l_{\eta} (x,k^2_{\perp}) + \sin \theta \, \psi^l_{\eta'} (x,k^2_{\perp})\,, \\
     \\
     \psi_{\eta_s}^s (x,k^2_{\perp}) = - \sin \theta \, \psi^s_{\eta} (x,k^2_{\perp}) + \cos \theta \, \psi^s_{\eta'} (x,k^2_{\perp})\,. \label{eq:LFWFmapping}
\end{array}
\end{eqnarray}
On the other hand, according to Eq.\,\eqref{eq:PDAdefinition}, the integration over $k_\perp$ yields:
\begin{eqnarray}
\begin{array}{cc}
     f_{\eta_l} \, \varphi_{\eta_l}^l (x) =  \cos \theta \, f^l_{\eta} \varphi^l_{\eta} (x) +  \sin \theta \, f^l_{\eta'} \varphi^l_{\eta'} (x)\,, \\
     \\
     f_{\eta_s} \, \varphi_{\eta_s}^s (x) = - \sin \theta \, f^s_{\eta} \varphi^s_{\eta} (x) + \cos \theta \, f^s_{\eta'} \varphi^s_{\eta'} (x)\,. 
\end{array}
\end{eqnarray}
Subsequently, using Eq.\,\eqref{eq:mapping2}, the $f_{\eta,\eta'}^{l,s}$ decay constants can be cast in terms of $f_{\eta_l,\eta_s}$, thus producing:
\begin{eqnarray}
\label{eq:RelDAs}
\begin{array}{cc}
\label{eq:DAmapping}
     \varphi_{\eta_l}^l (x) = \cos^2 \theta \, \varphi^l_{\eta} (x) + \sin^2 \theta \, \varphi^l_{\eta'} (x) \,,\\
     \\
     \varphi_{\eta_s}^s (x) = \sin^2 \theta \, \varphi^s_{\eta} (x) + \cos^2 \theta \, \varphi^s_{\eta'} (x)\,.
\end{array}
\end{eqnarray}
The above expressions are particularly useful. These explicitly allow the \emph{a priori} unknown $\eta_{l,s}$ DAs to be formulated in terms of the well-established $\varphi_{\eta,\eta'}^{l,s}$. The remaining task is to determine the masses of the constituent quarks $M_{u,s}$, of the $\eta_l$ and $\eta_s$ system, and compute the GPDs as well as the associated observables by employing the theoretical tools developed in this section. This procedure shall be addressed in the next section.

\section{Numerical results}
\label{sec:NumRes}

%%%%%
\begin{table*}[ht!]
\begin{tabular}[t]{l||c|c|c|c||c|c|c}
\hline
 & $f^l_{\eta}$ & $f^s_{\eta}$ & $f^l_{\eta'}$ & $f^s_{\eta'}$ & $f_{\eta_l}$ & $f_{\eta_s}$ & $\theta$ \\
\hline
Herein & \; 0.070 \; & \; -0.091 \; & \; 0.062 \; &\; 0.103\; &\; 0.093 \;&\; 0.137 \;& \;$41.50^\circ$\;\\
CSM & \; 0.074 \; & \; -0.092 \; & \; 0.068 \; & \; 0.101\; & \;0.101\; & \;0.138 \;& \;$42.95^\circ$\; \\
Lattice & \; 0.076(3) \; & \; -0.077(4) \; & \; 0.062(3) \; & \; 0.093(3) \; & \;0.098(3)\; & \;0.121(2) \;& \;$39.3(2.0)^\circ$\; \\
Phen. & \; 0.090(13) \; & \; -0.093(28) \; & \; 0.073(14) \; & \; 0.094(8)\; & \;0.116(11)\; & \;0.122(9) \;& \;$39.28(8)^\circ$\; \\
\hline
\end{tabular}
    \caption{$\eta-\eta'$ mixing angle and leptonic decay constants in the generic flavor basis and SA-MS, Eqs.\,\eqref{eq:chisl} and \eqref{eq:saBasis}, respectively. Results obtained from Continuum Schwinger Method (CSM)~\cite{Ding:2018xwy} and Lattice QCD~\cite{Ottnad:2025zxq} are shown for comparison. The  phenomenological average is derived from Refs.\,\cite{Feldmann:1998sh,DeFazio:2000my,Benayoun:1999au}.}
    \label{tab:decayconstants}
\end{table*}
%%%

\begin{figure}[!ht]
\begin{tabular}{c}
%    \centering
    \includegraphics[width=0.95\columnwidth ]{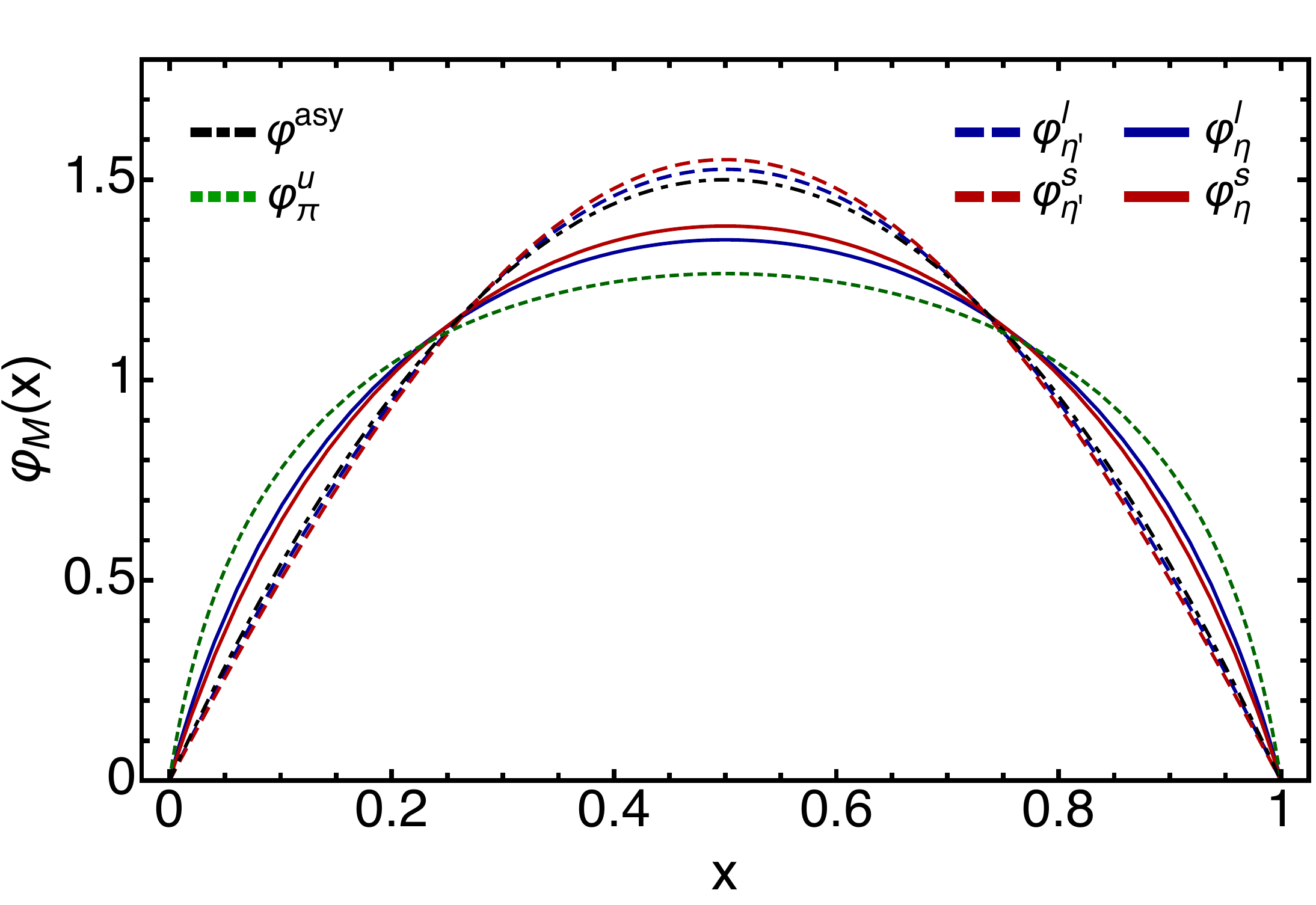} \\
    \includegraphics[width=0.95\columnwidth ]{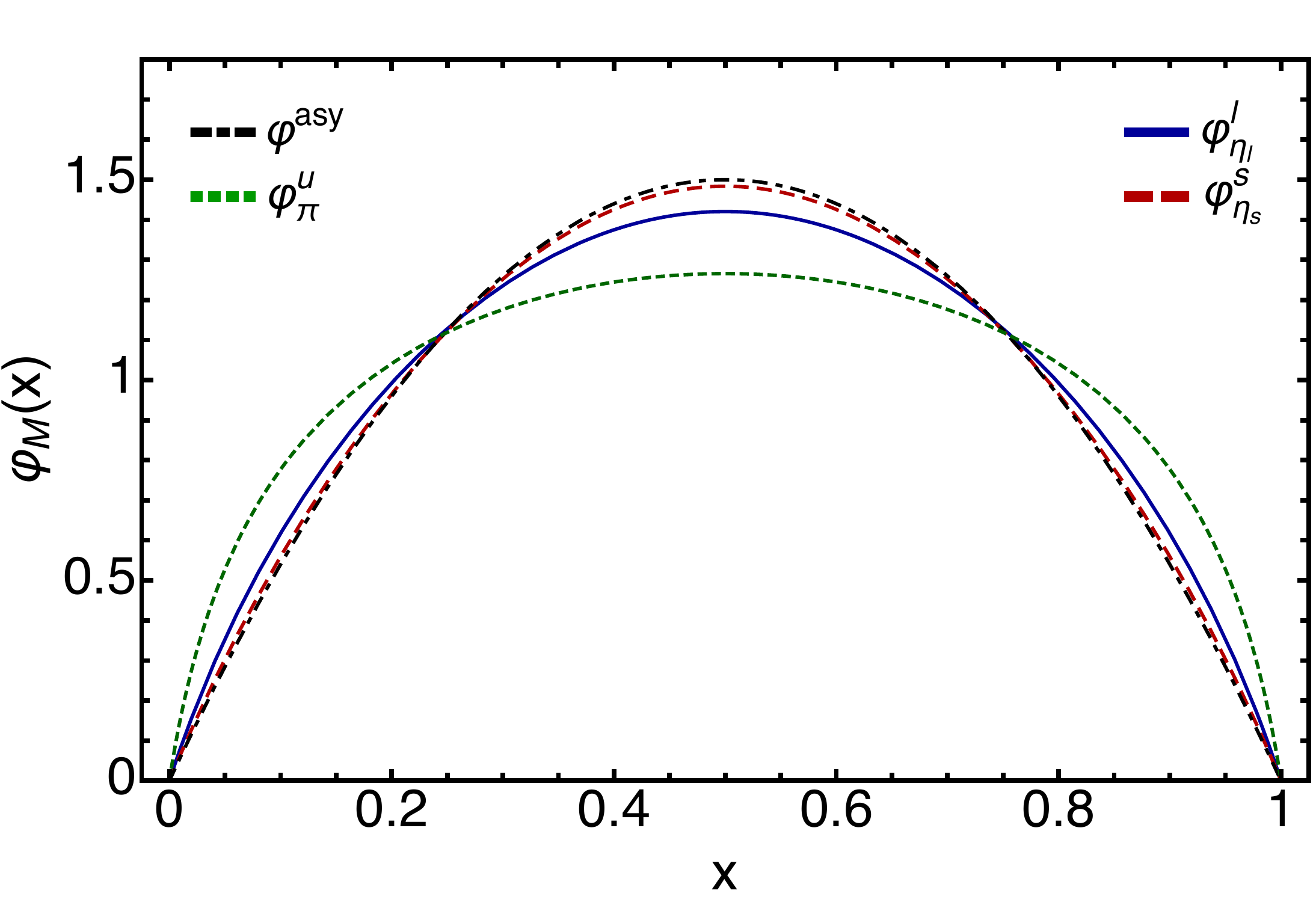}
\end{tabular}
    \caption{$\eta-\eta'$ related DAs. [\textbf{upper panel}] Those concerning the generic flavor basis which are employed as inputs for our analysis. [\textbf{lower panel}] The $\eta_{l,s}$ DAs are derived from Eq.\,\eqref{eq:RelDAs}. The pion DA, $\varphi_\pi^u$, and the asymptotic profile $\varphi^{\rm asy}$, Eq.\,\eqref{eq:DAasym}, are provided for the sake of comparison. }
    \label{fig:PDAs}
\end{figure}

We start by depicting the DAs related to the $\eta-\eta'$ system in Fig.\,\ref{fig:PDAs}. The inputs we have used, $\varphi_{\eta,\eta'}^{l,s}$, were first calculated in Ref.\,\cite{Ding:2018xwy} and  are conveniently parameterized as follows\,\cite{Raya:2024ejx}:
\begin{eqnarray}
\label{eq:PDAeta}
    \varphi_{\eta}^{l,s}(x;\zeta) &:=& n_{\rho}\,\ln \left[ 1 + \frac{x(1-x)}{(\rho_\eta^{l,s})^2}\right]\,,\\
    \varphi_{\eta'}^{l,s}(x;\zeta) &:=& n_{\rho'}\, x (1 - x)\,\text{exp}\left[\frac{x(1-x)}{(\rho_{\eta'}^{l,s})^2}\right]\,,
\end{eqnarray}
with $\rho_\eta^{l(s)} =0.329\,(0.421)$ and $\rho_{\eta'}^{l(s)}=1.7\,(1.221)$; $n_{\rho(\rho')}$ is the corresponding DA normalization constant. As can be readily observed in Fig.\,\ref{fig:PDAs}, the $\eta'$-related distributions are narrower than those of the $\eta$ meson.  Moreover, the former lie almost on top of the asymptotic distribution,
\begin{equation}
    \label{eq:DAasym}
    \varphi^{asy}(x):=6x(1-x)\,,
\end{equation}
whereas the ones of the $\eta$ meson resemble those of the pion more closely. In either case, the DAs associated with the $s$-quark tend to be slightly more compressed than the ones of their lighter-quark counterparts. The computed $\varphi^l_{\eta_l},\varphi^s_{\eta_s}$ profiles are marginally broader than the asymptotic limit; especially the latter exhibits marked similarity with the limiting result of asymptotic QCD.

%%%%%%%%
%%%%%%%%%

In order to compute $\varphi_{\eta_l}^{l}$ and $\varphi_{\eta_s}^{s}$ via Eq.\,\eqref{eq:DAmapping}, the mixing angle $\theta$ is required. We now address how to fix its value. We start by adopting the dressed quark masses from the pion and kaon analysis in Ref.\,\cite{Albino:2022gzs}, namely:
\begin{equation}
M_u=0.317\,\text{GeV}\,,\,M_s=0.574\,\text{GeV}\,.
\end{equation}
Subsequently, along with the mixing angle $\theta$, the values of $m_{\eta_l}$ and $m_{\eta_s}$ are set to provide sensible values for the corresponding leptonic decay constants. Therefore,
\begin{equation}
m_{\eta_l}=0.17\,\text{GeV}\,,\,m_{\eta_s}=0.84\,\text{GeV}\,,\theta=41.5^\circ\,,
\label{Fit parameters}
\end{equation}
where the resulting decay constants are listed in Table\,\ref{tab:decayconstants}. In particular, we find
\begin{equation}
    f_{\eta_l}\approx1.01 f_\pi\,,\,f_{\eta_{s}}\approx1.49 f_\pi\,,
\end{equation}
which shows a decent agreement with recent lattice QCD results~\cite{Ottnad:2025zxq}:
\begin{equation}
    f_{\eta_l}^{lat}\approx1.07(3) f_\pi\,,\,f_{\eta_{s}}^{lat}\approx1.31(3) f_\pi\,;
\end{equation}
and with the phenomenological average~\cite{Feldmann:1998sh,DeFazio:2000my,Benayoun:1999au}:
\begin{equation}
    f_{\eta_l}^{ph}\approx1.26(12) f_\pi\,,\,f_{\eta_{s}}^{ph}\approx1.33(10) f_\pi\,.
\end{equation}
Here $f_\pi=0.092$ GeV. A $5-10\%$ variation in the $\eta_l$ and $\eta_s$ masses barely affects the produced decay constants. In the typical octet-singlet basis\,\cite{Feldmann:1998sh,Feldmann:1998vh}, our computed values translate into:
\begin{eqnarray}
    f_8 := \sqrt{(f_{\eta_l}^2+2f_{\eta_s}^2)/3} &\approx&  1.35 \,f_\pi\,,\\
    f_1 := \sqrt{(2f_{\eta_l}^2+f_{\eta_s}^2)/3} &\approx&  1.2 \,f_\pi\,,
\end{eqnarray}
and the associated octet-singlet mixing angles, $\theta_{8[1]} = \theta-\arctan(\sqrt{2} f_{\eta_{s[l]}} / f_{\eta_{l[s]}})$, are found to be $\theta_8=-22.86^\circ$ and $\theta_1=-2.33^\circ$. For a comparison, the combined result of Refs.\,\cite{Feldmann:1998sh,DeFazio:2000my,Benayoun:1999au} yields:
\begin{eqnarray}
    f_8 = 1.34(8)\, f_\pi \,,& &f_1=1.25(10)\,f_\pi\,;\\
   \theta_8 = -18(6)^\circ \,,&&\theta_1=-6(6)^\circ\,.
\end{eqnarray}
One can also provide an estimate of the $\eta,\eta' \to \gamma \gamma$ decay widths via the phenomenological formulae from\,\cite{Feldmann:1997vc,Feldmann:1998vh}:
\begin{align}
\Gamma_{\eta{\rightarrow}\gamma\gamma}=&\frac{9\alpha^2_{em}}{64\pi^3}m^3_\eta\left[\frac{5}{9}\frac{f^l_\eta}{f_{\eta_l^{}}^2}+\frac{\sqrt{2}}{9}\frac{f^s_\eta}{f_{\eta_s}^2}\right]^2   \,,\\
\Gamma_{\eta^{\prime}{\rightarrow}\gamma\gamma}=&\frac{9\alpha^2_{em}}{64\pi^3}m^3_{\eta^{\prime}}\left[\frac{5}{9}\frac{f^l_{\eta^{\prime}}}{f_{\eta'_l}^2}+\frac{\sqrt{2}}{9}\frac{f^s_{\eta^{\prime}}}{f_{\eta'_s}^2}\right]^2 \,.
\end{align}
Based on the values presented in Table\,\ref{tab:decayconstants}, we find:
\begin{eqnarray}
    \Gamma_{\eta\to\gamma \gamma} = 0.52 \,\text{keV}\,,\,\Gamma_{\eta'\to\gamma \gamma} = 4.76 \,\text{keV}\,,
\end{eqnarray}
which are fairly compatible with the PDG average, see Ref.~\,\cite{ParticleDataGroup:2024cfk} ($0.515(18)\,\text{and}\,4.34(14)\,\text{keV}$, respectively), and with other explorations in Refs.\,\cite{Ding:2018xwy,Xu:2024frc,Zamora:2023fgl}. 
We would like to emphasize that these decay widths will be measured with higher precision in the 22 GeV upgrade of the JLab~\cite{Accardi:2023chb}. 
The related two-photon transition form factors shall be analyzed elsewhere, following the standard procedure\,\cite{Sultan:2024mva,Higuera-Angulo:2024oui}.

\begin{figure}[!ht]
%    \centering
    \includegraphics[width=0.90\columnwidth ]{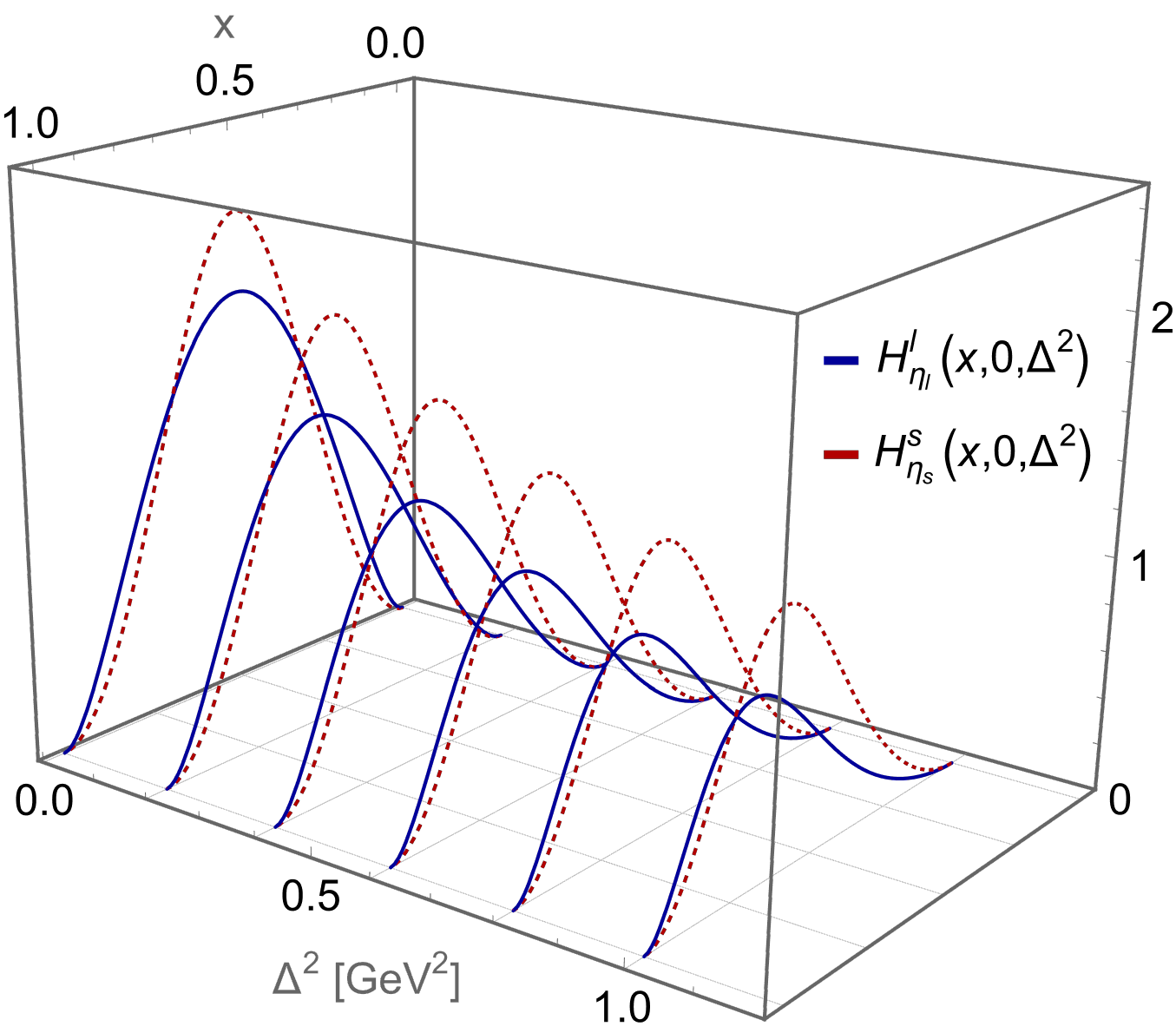}
    \caption{Valence-quark GPDs of the $\eta_l$ and $\eta_s$ system in the SA-MS, Eq.\,\eqref{eq:saBasis}. The curves show the $x$-dependence for different $\Delta^2$ values in the $\xi=0$ case.}
    \label{fig:GPDs}
\end{figure}

Provided $\varphi_{\eta_l}^{l}$ and $\varphi_{\eta_s}^{s}$, Eq.~\eqref{eq:DAmapping}, displayed in the lower panel of Fig.~\ref{fig:PDAs}, as well as the mixing angle $\theta$, fixed in Eq.~\eqref{Fit parameters}, it is now possible to compute the valence-quark skewless ($\xi=0$) GPDs for $\eta_l$ and $\eta_s$ by means of Eq.~\eqref{eq:defGPDgood}. The corresponding results, displayed in Fig.~\ref{fig:GPDs}, reveal a slightly milder $\Delta^2$ fall-off for the $\eta_s$ system as the momentum transfer increases.
Moreover, our results suggest that, in terms of dilation or compression effects, the corresponding GPDs follow a pattern similar to that of the resulting DFs, discussed below.

%%%%%%%
%%%%%%%

\begin{figure}[!ht]
%    \centering
    \includegraphics[width=0.95\columnwidth ]{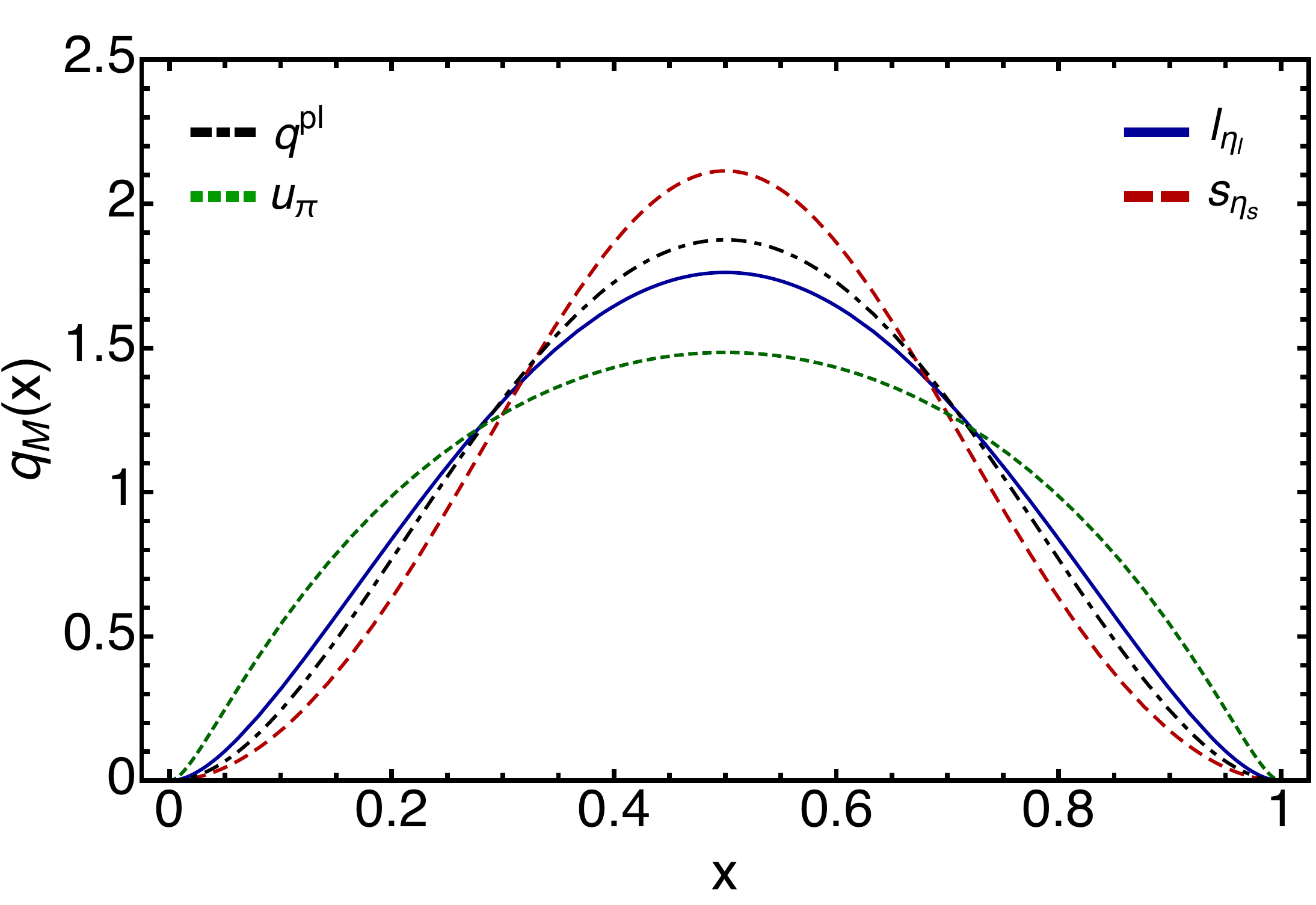}
    \caption{Valence-quark DFs of the $\eta_l$ and $\eta_s$ systems that characterize the $\eta-\eta'$ bound-states in the SA-MS, Eq.\,\eqref{eq:saBasis}. The DFs are defined at the hadronic scale $\zeta_H$. Pion DF and parton-like profiles, Eq.\,\eqref{eq:DFpl}, are shown for comparison.}
    \label{fig:PDFs}
\end{figure}

The corresponding DFs, which are derived through employing Eq.\,\eqref{eq:defGPDgood} and Eq.\,\eqref{eq:defPDF}, are depicted in Fig.\,\ref{fig:PDFs}. Unsurprisingly, these follow the same trends as the associated DAs: the DF of $\eta_s$ is narrower. In the same figure, we have also included the parton-like (pl) profile:
\begin{equation}
    \label{eq:DFpl}
    q^{\rm pl}(x):=30x^2(1-x)^2\,.
\end{equation}
This distribution emerges from squaring $\varphi^{asy}$, Eq.~\eqref{eq:DAasym}, and features no dilation or compression effects induced by the system's mass scales. Its position between the DFs of $\eta_l$ and $\eta_s$ is physically significant. It implies that the systems with $s\bar{s}$ components mark a boundary between the HF and the EHM mass generation dominance. We can also note that the pion DF is markedly broader than the rest of the distributions therein depicted. A final piece of observation is provided by the low-order Mellin moments listed in Table\,\ref{tab:momsDF}.
To some extent, the broadness of the DFs can be expected by noticing ($\xi_m=1-2x$):
\begin{eqnarray}
    \langle x^2 \rangle_{\pi} > \langle x^2 \rangle_{\eta_l} > &\langle x^2 \rangle_{\rm pl} = \frac{2}{7}& > \langle x^2 \rangle_{\eta_s}\,, \\
    \langle \xi_m^2 \rangle_{\pi} > \langle \xi_m^2 \rangle_{\eta_l} > &\langle \xi_m^2 \rangle_{\rm pl} = \frac{1}{7} &>\langle \xi_m^2 \rangle_{\eta_s} \,,
\end{eqnarray}
in particular, $\langle \xi_m^2 \rangle_{\eta_s}/\langle \xi_m^2 \rangle_{\pi} = 0.6$ already hints towards a notable compression of the $\eta_s$ DF with respect to that of the pion.

%%%%%%%%%
%%%%%%%%%%%%

%%%%%%%%%%%%%%%
\begin{table}[h!]
\begin{tabular}[t]{c|c|c|c|c|c|c}
\hline
& $\langle x^1 \rangle$ & $\langle x^2 \rangle$ & $\langle x^3 \rangle$ & $\langle x^4 \rangle$ & $\langle x^5 \rangle$ & $\langle (1-2x)^2 \rangle$ \\
\hline
$\pi$ &  \;$0.500$ \;&\; $0.300$ \;&\; $0.201$ \;&\; $0.143$ \;& \;$0.107$ \; & $0.200$\\
\;$\eta_l$ \;&  \;$0.500$ \;&\; $0.289$ \;&\; $0.183$ \;&\; $0.124$ \;& \;$0.088$ \;& $0.156$\\
\;$\eta_s$ \;&  \;$0.500$ \;&\; $0.280$ \;&\; $0.171$ \;&\; $0.111$ \;& \;$0.075$ \;& $0.120$\\
$q^{\rm pl}$ &  \;$0.500$ \;&\; $0.286$ \;&\; $0.176$ \;&\; $0.119$ \;& \;$0.083$ \;& $0.143$\\
\hline
\end{tabular}
\caption{Low-order Mellin moments of the $\pi$ and $\eta_{l,s}$ DFs at $\zeta_H$. The last row corresponds to the moments associated with the parton-like profile in Eq.\,\eqref{eq:DFpl}.}
\label{tab:momsDF}
\end{table}
%%%%%%%%%%%%%%%

%%%%%%%
\begin{figure}[!ht]
%    \centering
    \includegraphics[width=0.95\columnwidth ]{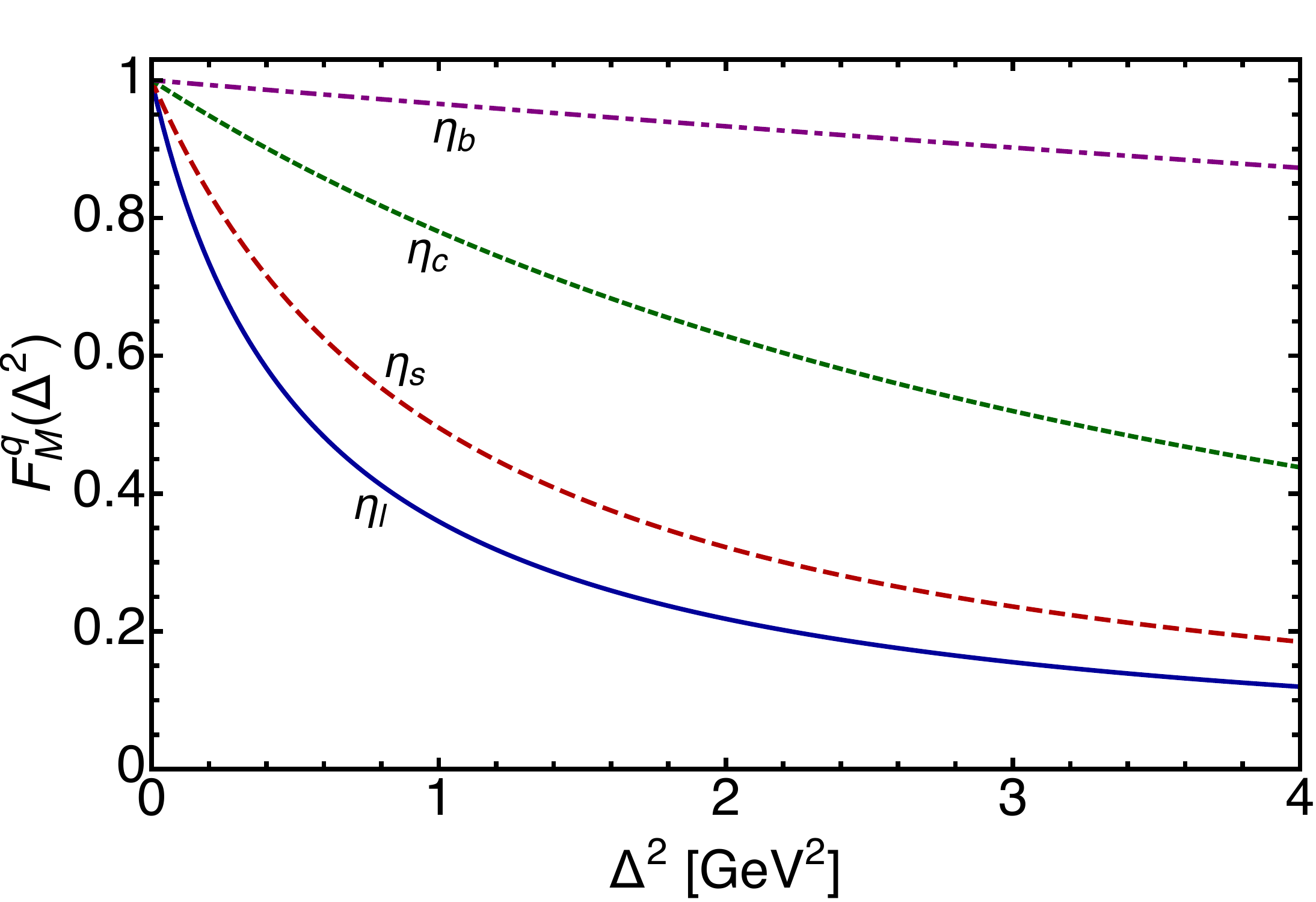}
    \caption{Electromagnetic FFs of the $\eta_l$ and $\eta_s$ system that characterize the $\eta-\eta'$ bound-states in the SA-MS, Eq.\,\eqref{eq:saBasis}. The analogous $\eta_c$ and $\eta_b$ profiles, Ref.\,\cite{Albino:2022gzs}, are also displayed.}
    \label{fig:EFFs}
\end{figure}
%%%%%%%

In addition, the electromagnetic FFs are computed by means of Eq.~\eqref{eq:EFFq} and displayed in Fig.\,\ref{fig:EFFs}, including the ones of  $\eta_c$ and $\eta_b$, which are taken from Ref.\,\cite{Albino:2022gzs}. It is clear that as the system becomes heavier, it exhibits a FF with a less pronounced $\Delta^2$ evolution, which would result in narrower charge distributions\,\cite{Raya:2024glv,Xu:2024vkn}. The associated charge radii are listed in Table\,\ref{tab:radii}. This quantity provides an estimate of the hadron's size and,  understandably, it turns out to be inversely related to the bound-state's mass. Notably,  the product $f_{\M}\cdot r_{\M}$ remains nearly constant up to the charm quark sector, as expected from QCD-driven backgrounds\,\cite{Chen:2018rwz}.

%%%%%%%
\begin{table}
\begin{tabular}[t]{c||c|c|c}
\hline
 & $m_{\M}$ [GeV] & $r_{\M}$ 
 [fm] & $f_{\M}\cdot r_{\M}$  \\
\hline
$\pi$ &  0.14 & 0.659 & 0.307 \\
$\eta_l$ &  0.17 & 0.655 & 0.308 \\
%$K$ &  0.49 & 0.600 & 0.335 \\
$\eta_s$ &  0.84 & 0.472 & 0.330 \\
$\eta_c$ &  2.98 & 0.255 & 0.301 \\
$\eta_b$ &  9.39 & 0.088 & 0.217 \\
\hline
\end{tabular}
\caption{Masses, charge radii, and the product $f_{\M}\cdot r_{\M}$. With the exception of $\eta_l$ and $\eta_s$, all other results are taken from Ref.\,\cite{Albino:2022gzs}, which employs the same framework.}
\label{tab:radii}
\end{table}
%%%%%

\begin{figure}[!ht]
%    \centering
    \includegraphics[width=0.90\columnwidth ]{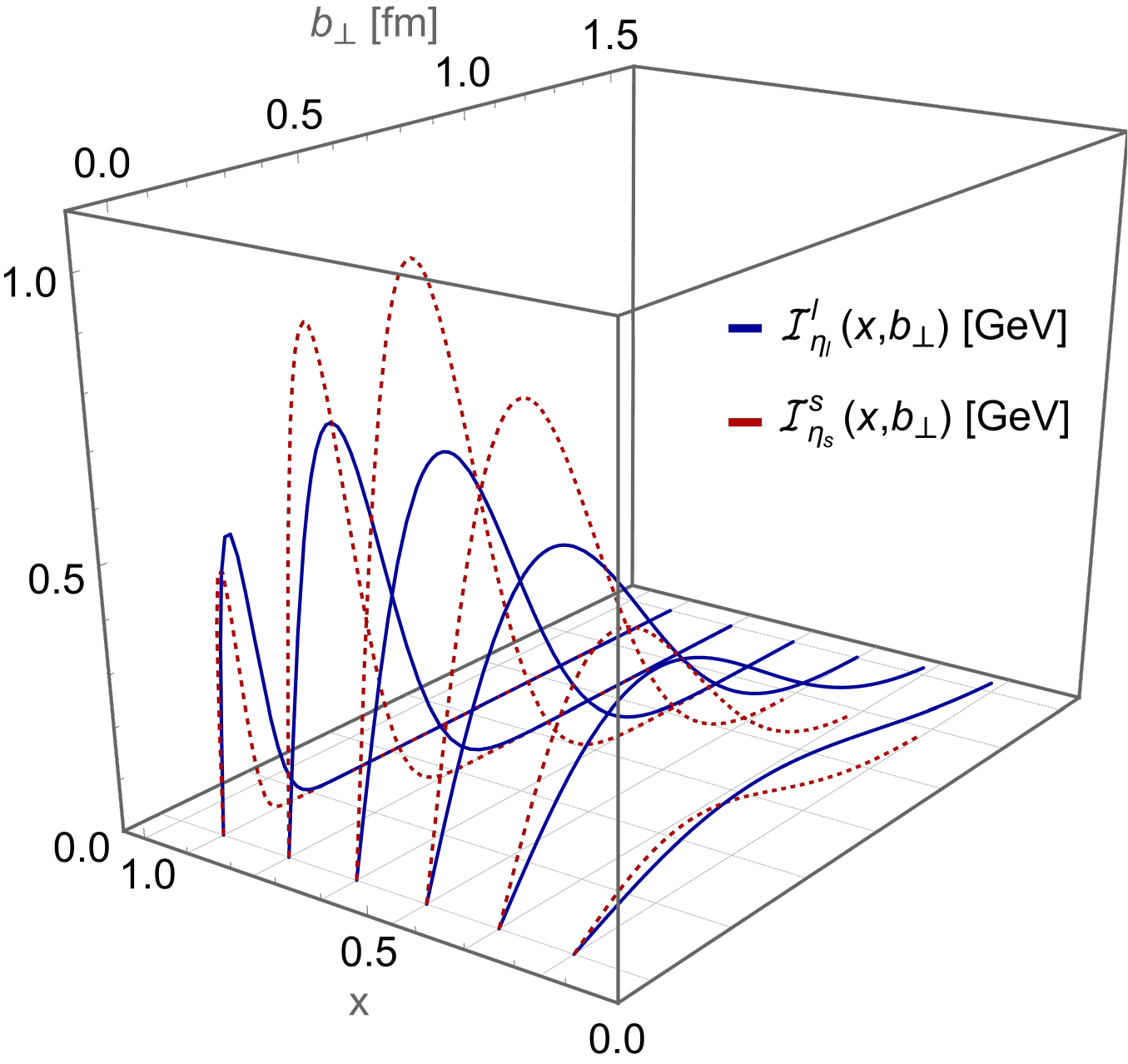}
    \caption{The valence-quark IPS-GPDs of the $\eta_l$ and $\eta_s$ system in the SA-MS, Eq.\,\eqref{eq:IPSdef}. The different curves capture the $|b_\perp|$-dependence for several values of $x$.}
    \label{fig:IPS}
\end{figure}

Finally, Eq.~\eqref{eq:IPSdef} allows us to compute the corresponding IPS-GPD, shown in Fig.~\ref{fig:IPS}. As a function of $|b_\perp|$, the distribution for the heavier $\eta_s$ system appears noticeably more compressed while the one corresponding to $\eta_l$ shows a less pronounced falloff. Intuitively, this suggests that the heavier system is likely to be significantly more localized in its spatial extent. The location ($x^{\text{max}}$,$|b_\perp|^{\text{max}}$) and magnitude ($\mathcal{I}_{\M}^{q-\text{max}}$) of the maximal values are collected in Table~\ref{tab:ImpParMax}. 
%%%%%%%
%%%%%%%
%%%%%%
\begin{table}[ht]
\centering
\begin{tabular}[t]{l||c|c|c}
\hline
 & $x^{\text{max}}$ & $|b_\perp|^{\text{max}}\, \text{[fm]}$ & $\mathcal{I}_{\M}^{q-\text{max}}\,\text{[GeV]}$  \\
\hline
$\pi$ & $0.90$ & $0.07$ &  $0.94$ \\
$\eta_l$   & $0.70$ & $0.22$ &  $0.74$ \\
$\eta_s$ & $0.63$ & $0.20$ &  $1.04$ \\
$\eta_c$ & $0.53$ & $0.14$ &  $3.09$  \\
$\eta_b$ & $0.51$ & $0.05$ &  $10.97$  \\
\hline
\end{tabular}
\caption{Location ($x^{\text{max}}$,$|b_\perp|^{\text{max}}$) and magnitude ($\mathcal{I}_{\M}^{q-\text{max}}$) of the global maximum for the IPS-GPDs of the $\eta-\eta'$ system. We have included the pion and heavy-quarkonia results from Ref.\,\cite{Albino:2022gzs}.}
\label{tab:ImpParMax}
\end{table}

The values presented in Table~\ref{tab:ImpParMax} follow a consistent trend:  as the meson mass increases, the distributions exhibit progressively larger maxima. Moreover, it is clear that $x^{\text{max}}$ tends to approach $1/2$ as the system becomes infinitely heavy\,\cite{Raya:2024ejx}. As far as $|b_\perp|$ is concerned, for any system with isospin symmetry, its expectation value is\,\cite{Raya:2021zrz}:
\begin{equation}
    \langle |b_\perp|^2\rangle_{\M}^q = \frac{2}{3} r_{\M}^2\,;
\end{equation}
consequently, as the hadron mass increases, its size would diminish, ultimately approaching the point particle limit.

\section{Summary}
\label{sec:Summary}

We have employed a form-invariant algebraic model to describe the mixed $\eta-\eta'$ system. This framework has been successfully applied to ground-state pseudoscalars\,\cite{Albino:2022gzs,Almeida-Zamora:2023bqb}, as well as ground-state vector mesons\,\cite{Almeida-Zamora:2023rwg}. Its extension to the $\eta-\eta'$ mesons, presented in this work, yields a physically congruent picture.

Firstly, the obtained mixing angle and decay constants are found to be fully compatible with previous results,\,\cite{Ding:2018xwy}. Moreover, the resulting DAs of the physical states exhibit the expected pattern: they reveal an asymptotic-like profile, with those corresponding to the strange quark being narrower. As has already been established, DAs for pions and kaons are broad, while those for states with at least one heavy quark are more compressed. Therefore, this proximity to the asymptotic profile highlights that systems with $s\bar{s}$ components mark a boundary between strong and weak mass generation mechanisms\,\cite{Raya:2024ejx}. 

The DFs of the valence quarks follow a similar pattern. The one corresponding to the $\eta_l$ state is broader than that of $\eta_s$, while the parton-like profile lies in between these curves. This behavior is attributed to the emergence of an $x-k_\perp$ correlation in their respective LFWFs, which in turn is influenced by the mass of the state in question. Unsurprisingly, the associated GPDs show  similar compression/dilation patterns, with the $\eta_s$ system displaying a less pronounced $\Delta^2$ evolution.

A complementary perspective comes from electromagnetic FFs, where heavier systems exhibit a slower decay. This behavior reflects an inverse relationship with the mass of the system and its decay constant, which corresponds to its smaller spatial extent\,\cite{Raya:2024glv}. Remarkably, the product $f_{\M}\cdot r_{\M}$ seems to remain constant up to the mass sector of the charm quark.

Another perspective on the spatial structure is provided by the IPS-GPDs, demonstrating once again that heavier states are more spatially compressed but reaching higher peaks at smaller transverse separations. Moreover, as the meson mass increases, the current model is fully capable of capturing the anticipated point-like limiting behavior.
Thus, based on the quantities analyzed here, the extension of the algebraic model to the mixed $\eta-\eta'$ system yields remarkably robust results. Moving forward, the aim is to examine complementary physical quantities of interest, including the $\eta,\eta' \to \gamma^{*}\gamma^{(*)}$ transition form factors, in addition to the corresponding excited states.

%%%%%%%%%%%%%%%%%%%%%%%%%%%%%%%%%%%%%%%%%%%%%%%%%%%%%%%%%%%
%%%%%%%%%%%%%%%%%%%%%%%%%%%%%%%%%%%%%%%%%%%%%%%%%%%%%%%%%%%
%%%%%%%%%%%%%%%%%     Algebraic Model     %%%%%%%%%%%%%%%%%%%%%
%%%%%%%%%%%%%%%%%%%%%%%%%%%%%%%%%%%%%%%%%%%%%%%%%%%%%%%%%%%
%%%%%%%%%%%%%%%%%%%%%%%%%%%%%%%%%%%%%%%%%%%%%%%%%%%%%%%%%%%

%%%%%%%%%%%%%%%%%%%%%%%%%%%%%%%%%%%%%%%%%%%%%%%%%%%%%%%%%%%%%%%
%%%%%%%%%%%%%%%%%%%%%%%%%%%%%%%%%%%%%%%%%%%%%%%%%%%%%%%%%%%%%%%
%%%%%%%%%%%%%%%%%%%%     BIBLIOGRAPHY      %%%%%%%%%%%%%%%%%%%%
%%%%%%%%%%%%%%%%%%%%%%%%%%%%%%%%%%%%%%%%%%%%%%%%%%%%%%%%%%%%%%%
%%%%%%%%%%%%%%%%%%%%%%%%%%%%%%%%%%%%%%%%%%%%%%%%%%%%%%%%%%%%%%%

\section*{Acknowledgment}
This work is supported by the Spanish MICINN grant PID2022-140440NB-C22, and the regional Andalusian project P18-FR-5057.
L.A. acknowledges financial support provided by CONAHCyT (Mexico) Project No. 320856 ({\em Paradigmas y Controversias de la Ciencia 2022}), as well as the support of Ayuda B3 {\em ``Ayudas para el desarrollo de l\'\i neas de investigaci\'on propias" del V Plan Propio de Investigaci\'on y Transferencia 2018-2020}, University of Pablo de Olavide, Seville, Spain.
The work of R.J.H.P is partly supported by SECIHTI (Mexico) through projects No. 320856 (Paradigmas y Controversias de la Ciencia 2022), CBF2023-2024-268 and {\em Sistema Nacional de Investigadores}.
A.B. acknowledges the support of the {\em Coordinaci\'on de la Investigaci\'on Cient\'ifica} of the{\em Universidad Michoacana de San Nicol\'as de Hidalgo}, Morelia, Mexico,  grant no. 4.10, {\em Secretaría de Ciencia, Humanidades, Tecnología e Inovación} (SECIHTI) (Mexico), project CBF2023-2024-3544
as well as the Beatriz-Galindo support during his scientific stay at the University of Huelva, Huelva, Spain.
B.A.Z. acknowledges CONACyT (CVU 935777) (Mexico) for PhD fellowship.

\bibliography{References}

\end{document}